\documentclass{PoS}
\pdfoutput=1
\usepackage[utf8]{inputenc}
\usepackage{cite}
\usepackage[hyphens]{url} 
\usepackage{lmodern}
\usepackage[safe]{textcomp}
\usepackage{lmodern} 
\usepackage{bbm} 
\usepackage{amstext}  
\usepackage{amsfonts}
\usepackage{amsthm}
\usepackage{graphicx}
\usepackage{amsmath}
\usepackage{amsfonts}
\usepackage{mathtools} 
\usepackage{array}                
\usepackage{xcolor}  
\usepackage[absolute]{textpos}
\usepackage{multirow} 
\usepackage{slashed}
\usepackage{relsize}

\newcommand{\OpenLoops}{\text{\sc OpenLoops}}
\newcommand{\Collier}{\text{\sc Collier}}
\newcommand{\Cuttools}{\text{\sc Cuttools}}

\newcommand{\f}[2]{\frac{#1}{#2}}

\newcommand{\ssst}[1]{\scriptscriptstyle{\text{#1}}}
\newcommand{\nosss}[1]{#1}

\newcommand{\bit}{\begin{itemize}}
\newcommand{\eit}{\end{itemize}}
\newcommand{\bce}{\begin{center}}
\newcommand{\ece}{\end{center}}
\newcommand{\bea}{\begin{eqnarray}}
\newcommand{\eea}{\end{eqnarray}}
\newcommand{\be}{\begin{equation}}
\newcommand{\ee}{\end{equation}}
\newcommand{\ba}{\begin{align}}
\newcommand{\ea}{\end{align}}
\newcommand{\beas}{\begin{eqnarray*}}
\newcommand{\eeas}{\end{eqnarray*}}
\newcommand{\bes}{\begin{equation*}}
\newcommand{\ees}{\end{equation*}}
\newcommand{\bas}{\begin{align*}}
\newcommand{\eas}{\end{align*}}

\newcommand{\eps}{{\varepsilon}}

\newcommand{\lb}{\left(}
\newcommand{\rb}{\right)}

\newcommand{\idop}{1\!\!\!1}

\newcommand{\Dbar}[1]{\bar{D}_{\nosss{#1}}}

\newcommand{\momp}[1]{p_{#1}}
\newcommand{\mass}[1]{m_{\nosss{#1}}}
\newcommand{\heli}{h}

\newcommand{\momq}{\bar{q}}
\newcommand{\tilq}{\tilde{q}}

\newcommand{\calA}{\mathcal{A}}
\newcommand{\calC}{\mathcal{C}}

\newcommand{\calM}{\mathcal{M}}
\newcommand{\calN}{\mathcal{N}}

\newcommand{\calU}{\mathcal{U}}

\newcommand{\calW}{\mathcal{W}}

\newcommand{\seg}{S}

\newcommand{\col}{\mathrm{col}}
\newcommand{\hel}{\mathrm{hel}}
\newcommand{\re}{\mathrm{Re}}
\newcommand{\Tr}{\mathrm{Tr}}

\newcommand{\rd}{\mathrm d}

\definecolor{bluemar}{rgb}{0,0,.5}
\definecolor{redmar}{rgb}{.8,0,0}
\definecolor{greenmar}{rgb}{0,.5,0}

\title{A new method for one-loop amplitude generation and reduction in OpenLoops}

\ShortTitle{A new method for one-loop amplitude generation and reduction in OpenLoops}

\author{Federico Buccioni\\
        University of Zurich, Zurich, SWITZERLAND\\
        E-mail: \email{buccioni@physik.uzh.ch}}
        
\author{Stefano Pozzorini\\
        University of Zurich, Zurich, SWITZERLAND\\
        E-mail: \email{pozzorin@physik.uzh.ch}}
        
\author{\speaker{Max Zoller}\\
       University of Zurich, Zurich, SWITZERLAND\\
       E-mail: \email{zoller@physik.uzh.ch}}       

\abstract{We describe a new method \cite{Buccioni:2017yxi} for the automated construction of one-loop amplitudes based on the open-loop algorithm,
where various operations are performed on-the-fly while constructing the integrand. 
In particular, an on-the-fly reduction interleaved
with the construction steps of the amplitude keeps the maximum tensor rank  
in the loop momentum at two throughout the algorithm, thus drastically reducing the complexity of the calculation.
The full reduction to scalar integrals is unified with the amplitude construction in a single recursion
within the \OpenLoops{} framework.
This approach strongly exploits the factorisation of one-loop integrands
in a product of loop segments.
The on-the-fly approach, which is also applied to helicity summation and 
the  merging of different diagrams, increases the speed of the original open-loop algorithm in a very significant way.  
A remarkably high level of numerical stability is achieved by exploiting freedoms in reduction identities and through
simple expansions in rank-two Gram determinants. These features are particularly attractive for NLO multi-leg
and NNLO real--virtual calculations.
The new algorithm will be made public in a forthcoming release of the {\sc OpenLoops} program.
}

\FullConference{13th International Symposium on Radiative Corrections (Applications of Quantum Field Theory to Phenomenology)\\
         25-29 September, 2017 \\
         St. Gilgen, Austria}

\begin{document}

\section{Automated amplitude generation in OpenLoops{}}
\begin{figure}[t!]\begin{center} \begin{tabular}{ccc}
\includegraphics[height=45mm]{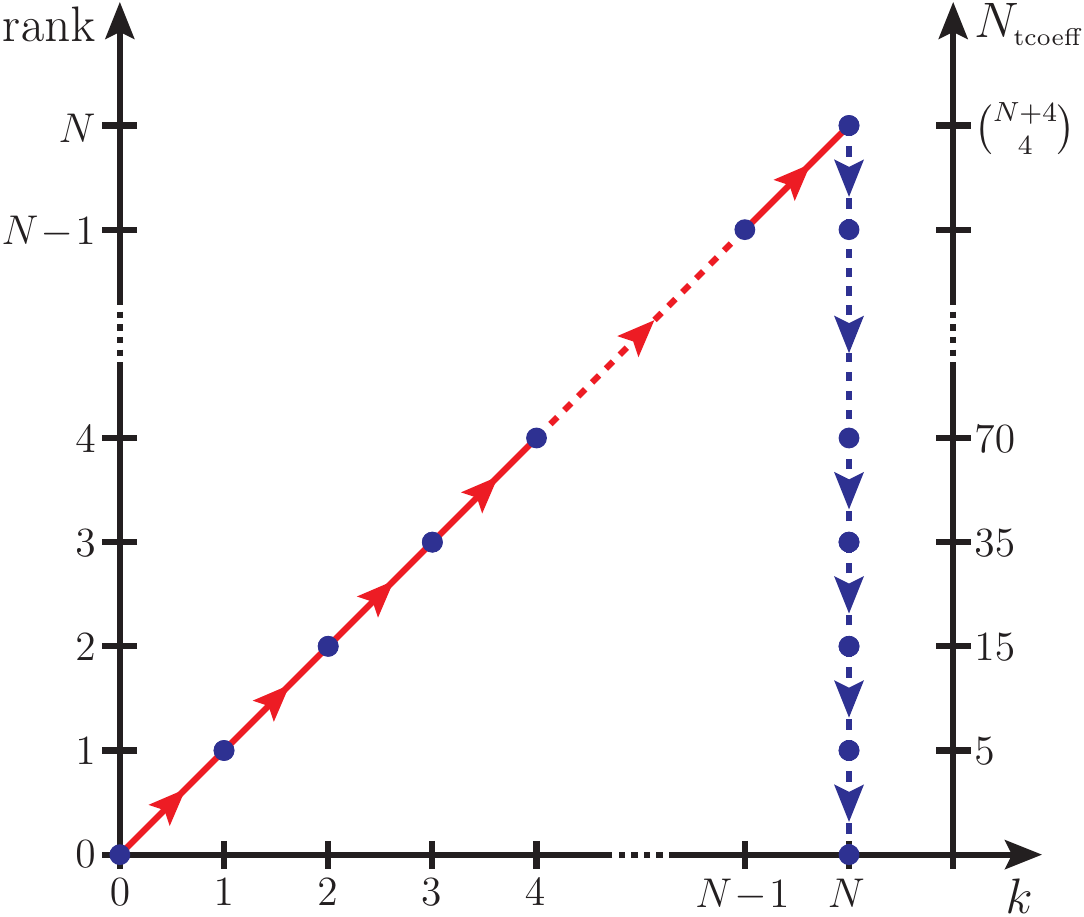} & \quad\; &
\includegraphics[height=45mm]{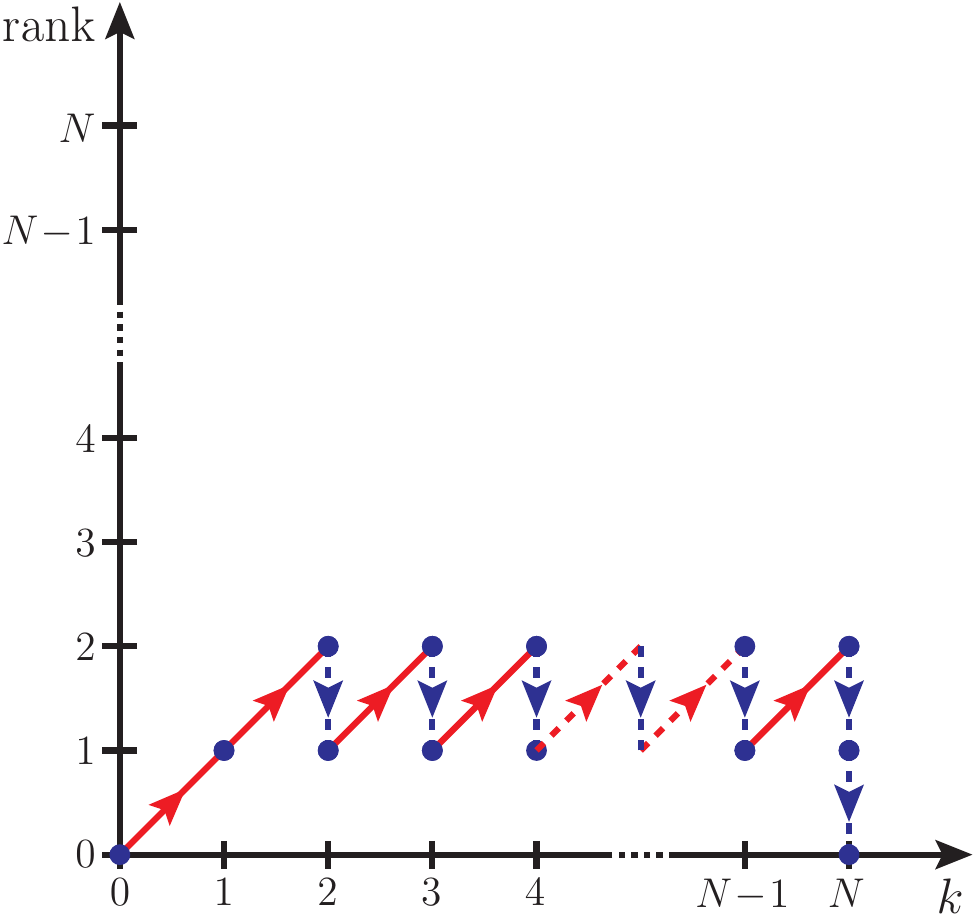}\\ (a) A posteriori reduction & & (b) On-the-fly reduction
 \end{tabular} \end{center}\vspace{-2mm}
\caption{Evolution of the tensor rank $R$ and number
$N_{\mathrm{tcoeff}}\,(R)=\binom{R+4}{4}$ of open-loop tensor coefficients
with the number $k$ of dressed segments (if each dressing step increases
the rank by $1$). \label{fig:OL1OL2_r_vs_n} } 
\end{figure}

\noindent
In the last decade, powerful methods for the calculation of one-loop scattering amplitudes have been developed. 
Highly automated one-loop tools, such as \OpenLoops \cite{Cascioli:2011va,hepforge},
provide the key to achieving NLO precision in multi-purpose Monte Carlo generators. 
The helicity- and colour-summed scattering probability densities for an $n$-particle process
\bea
\calW_{\ssst{LO}}=
\sum\limits_{\hel,\col}
|\calM_{0}|^2,\quad
\calW_{\ssst{NLO}}^{\ssst{virtual}}=
\sum\limits_{\hel,\col} 
2\,\re \Big[\calM_{0}^*\calM_{1}\Big] \quad \text{with}\;
\calM_{l}=\sum\limits_d\calM_{l}^{(d)}
%
\label{M2W}
\eea
are computed as sums of Feynman diagrams $d$ with $l=0,1$ loops and $n$ external particles.
A one-loop Feynman diagram amplitude can be written as
\bea \mathcal{M}^{(d)}_{1}&=& \mathcal{C}^{(d)}_1\,
\int\!\rd\momq\, \f{\Tr \Big[{\calN}(q)\Big]}{\Dbar{0}(\momq)\cdots \Dbar{N-1}(\momq)}
\,= \,
\vcenter{\hbox{\scalebox{1.}{\includegraphics[height=20mm]{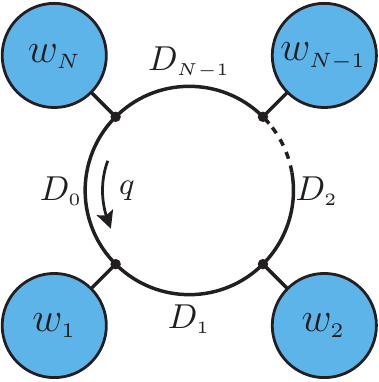}}}}
{}\label{eq:A1d} \eea
with a colour factor $\mathcal{C}^{(d)}_1$ and scalar propagators $\Dbar{i}(\momq)=(\momq + \momp{i})^2-\mass{i}^2$, which contain the loop momentum $\momq$,
the mass $\mass{i}$ and the external momentum $\momp{i}$.\footnote{The bar marks $D$-dimensional quantities as opposed to four-dimensional ones, where $D=4-2\eps$.} 
This is computed by cutting the loop open at one propagator $\Dbar{0}$ and numerically constructing the numerator\footnote{Here we explicitly include the dependence on the global helicity configuration $\heli$,
or equivalently the helicity configurations $\heli_1,\ldots,\heli_N$ of the external subtrees in the individual segments.} 
\be
\Big[\calN(q,\heli)\Big]_{\beta_0}^{\beta_N}= \vcenter{\hbox{\includegraphics[height=20mm]{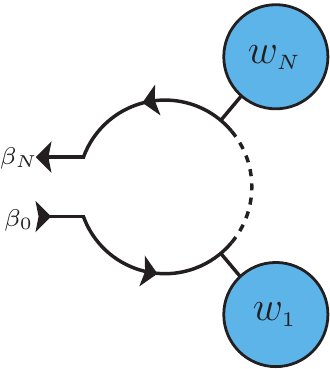}}}=
\Big[\seg_1(q,\heli_1)\Big]_{\beta_0}^{\beta_1}\,
\Big[\seg_2(q,\heli_2)\Big]_{\beta_1}^{\beta_2}\cdots
\Big[\seg_{N}(q,\heli_{N})\Big]_{\beta_{N-1}}^{\beta_N}, \label{eq:fac}
\ee
where $\beta_{0,N}$ are the Lorentz or spinor indices of the cut propagator. Finally,
the trace is taken by multiplying with $\delta_{\beta_0}^{\beta_N}$ and the tensor integrals are evaluated.
A key feature exploited in our approach is the factorisation of the numerator into segments $\seg_i(q,\heli_i)$, 
each consisting of a loop vertex and propagator and one or two external subtrees $w_i(\heli_i)$,
\be
\Big[\seg_{i}(q,\heli_i)\Big]_{\beta_{i-1}}^{\beta_{i}}
=\quad
\raisebox{3mm}{\parbox{12mm}{
\includegraphics[height=14mm]{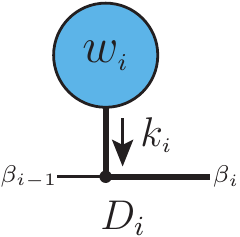}}} \qquad \text{or} \qquad
\Big[\seg_{i}(q,\heli_i)\Big]_{\beta_{i-1}}^{\beta_{i}}
=\quad
\raisebox{3mm}{\parbox{12mm}{
\includegraphics[height=14mm]{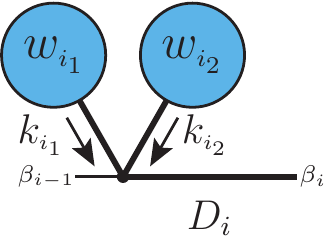}}}{}\quad\quad\;{}.
\label{eq:seg3point}
\ee
We construct the numerator recursively through so-called {\it dressing} steps
\be
\calN_k(q)=\calN_{k-1}(q)\seg_k(q), \qquad k \leq N{},
\label{eq:OLrec}
\ee
starting from the initial condition $\calN_{0}=\idop$.
The partially dressed numerator, which we call an open loop, is a $q$-polynomial\footnote{
Dressing steps are performed numerically at the level of the coefficients $\calN^{(n)}_{\mu_1\dots\mu_r}$},
\bea \calN_{k}(q) &=&\raisebox{3mm}{\parbox{65mm}{
\includegraphics[height=14mm]{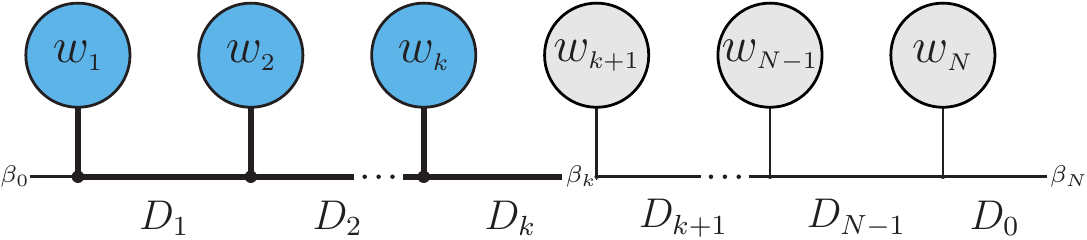}}} =\prod\limits_{i=1}^{k}\seg_i(q)
= \sum\limits_{r=0}^R
\calN^{(k)}_{\mu_1\dots\mu_r}
\,q^{\mu_1}\cdots q^{\mu_r}.\qquad\\[-2mm]
& &\underbrace{\phantom{xxxxxxxxxxxxxxx}}_{\text{dressed segments}}\;
 \underbrace{\phantom{xxxxxxxxxxxxx}}_{\text{undressed segments}}\nonumber
 \eea
The rank $R$ is potentially increased by one in each step (see Fig.~\ref{fig:OL1OL2_r_vs_n} (a)).
In the first version of \OpenLoops{} \cite{Cascioli:2011va,hepforge} the tensor reduction was performed a posteriori with external libraries, such as  \Collier{} \cite{Denner:2016kdg} or \Cuttools{} \cite{Ossola:2007ax}.
The basic idea of the on-the-fly reduction is to perform reduction steps at integrand level interleaved with the dressing steps (see Fig.~\ref{fig:OL1OL2_r_vs_n} (b)), keeping the rank and hence the complexity
of the tensor coefficients low throughout the calculation.
\section{The on-the-fly method}
\begin{figure}[t!]\begin{center}\begin{minipage}{0.5\textwidth}\begin{center}
                             \includegraphics[width=\textwidth]{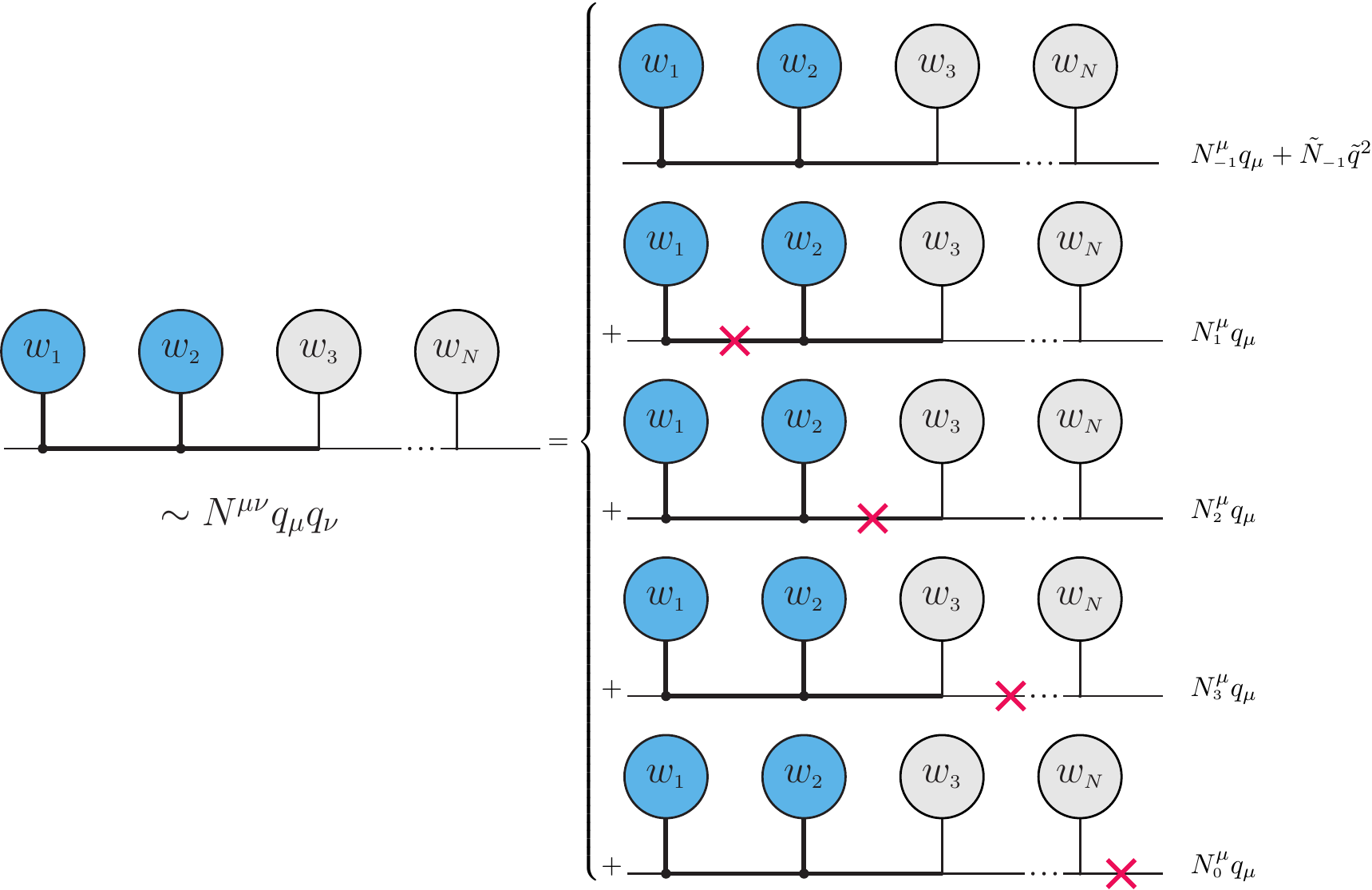}\\(a) \end{center}
                            \end{minipage}\qquad
\begin{minipage}{0.4\textwidth}\begin{center}
 \includegraphics[width=0.6\textwidth]{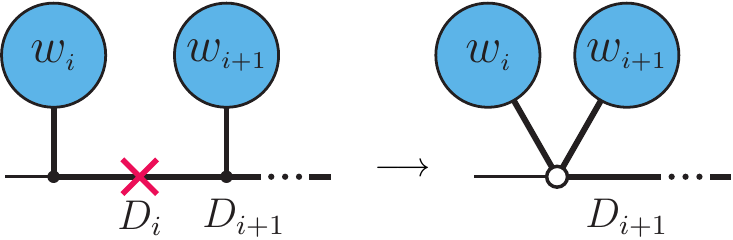}\\(b)\\[5mm] \includegraphics[width=\textwidth]{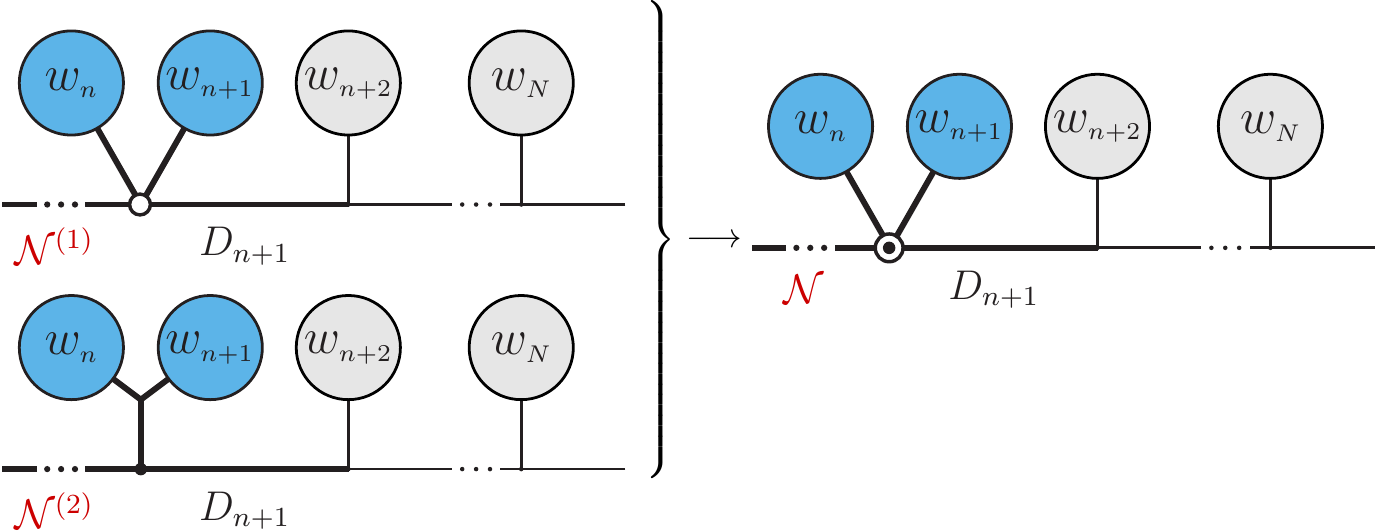}\\(c) \end{center}
\end{minipage}
              \end{center}\vspace{-2mm}
 \caption{Creation of pinched topologies in on-the-fly reduction steps (a).
 Two dressed segments with a pinched propagator in between can be treated as one effective segment (b) of an $(N-1)$-point open loop. In many cases such
 an open loop can be absorbed into another $(N-1)$-point open loop stemming from a genuine $(N-1)$-point Feynman diagram (c).
 \label{fig:topos}}
\end{figure}

\noindent
The on-the-fly reduction formulas are based on \cite{delAguila:2004nf} and have the form\footnote{This is also valid for triangles in renormalisable theories \cite{delAguila:2004nf,Buccioni:2017yxi} if we set terms
involving $\Dbar{3},\momp{3}$ to zero.}
\bea
q^\mu q^\nu  &=& =  \sum\limits_{i=-1}^{3}\lb A^{\mu\nu}_{i} + B^{\mu\nu}_{i,\lambda}\,q^{\lambda} \rb D_i(q) {}, \label{eq:qqred}
 \qquad D_i(q)=\begin{cases}
             1, & i=-1\\
             (q + \momp{i})^2-\mass{i}^2, & i\geq 0  
             \end{cases} 
\eea
where the coefficients $A^{\mu\nu}_{i}$ and $B^{\mu\nu}_{i,\lambda}$ are $q$-independent. The loop momentum dependence resides in the reconstructed denominators $D_i=\Dbar{i}-\tilq^2$ which
cancel denominators in the full integrand.\footnote{The terms $\propto \tilq^2$, 
where $\tilq=\momq-q$ is $(D-4)$-dimensional, lead to rational terms of type $R_1$ \cite{delAguila:2004nf}.}
Hence in each reduction step from rank-2 to rank-1 open loops new topologies with pinched propagators are created. 
Due to the factorisation \eqref{eq:fac} of the open loop \eqref{eq:qqred} can be applied
to a partial integrand, e.g.~after two dressing steps (see Fig.~\ref{fig:topos} (a)),
irrespective of segments to be dressed in subsequent steps and further $\Dbar{i}$,
\be \left[\f{\calN^{\mu\nu}q_\mu q_\nu}{\Dbar{0}\Dbar{1}\Dbar{2}\Dbar{3}} \right] \prod\limits_{i=k+1}^{N}\f{\seg_i(q)}{\Dbar{i-1}}
=\left[\f{\calN^{\mu}_{-1}q_\mu +\calN_{-1}+\tilde{\calN}_{-1}\tilq^2}{\Dbar{0}\Dbar{1}\Dbar{2}\Dbar{3}} 
+\sum\limits_{i=0}^{3}\f{\calN^{\mu}_{i}q_\mu +\calN_{i}}{\Dbar{0}\cdots\slashed{\Dbar{i}}\cdots\Dbar{3}}\right]
\prod\limits_{i=k+1}^{N}\f{\seg_i(q)}{\Dbar{i-1}}.
\ee
The reduction step is then followed by the next dressing steps and further reduction steps for each resulting topology until the loop is fully dressed and final reduction steps
are performed at integral level (for details see \cite{Buccioni:2017yxi}). The freedom to choose $\Dbar{0},\ldots,\Dbar{3}$ from the set of available scalar propagators allows us to avoid numerical instabilities due to vanishing Gram determinants in four- and higher-point
integrands and effectively shift them to triangles. Here we identified a simple kinematical configuration responsible for numerical instabilities due to a vanishing rank-two Gram determinant $\Delta$
and perform an analytical expansion in $\Delta$ \cite{Buccioni:2017yxi}. In this way we completely avoid Gram determinant instabilities in our algorithm,
which leads to excellent numerical stability (see section~\ref{sec:numstab}).\\
An important issue is the proliferation of pinched topologies, especially with many subsequent on-the-fly reductions steps. The solution we implement is an on-the-fly merging of open loops,
which have the same topology\footnote{A topology is defined as an ordered set $\{\Dbar{0},\ldots,\Dbar{N-1}\}$ of propagator denominators.}, 
and the same segments to be dressed in subsequent steps.
This means that we sum the $\beta_0$, $\beta_k$, helicity and $q$-tensor components of the open loops to be merged into a single open loop. All future dressing and reduction steps are then performed
on the merged open loop which in general contains a set of contributions from different Feynman diagrams. 
In this way we can absorb most of the pinched contributions into lower-point Feynman diagrams (see Fig.~\ref{fig:topos} (c)) with the same topology and undressed segments.
A requirement is that the segments left and right of the pinched propagator are already dressed, in which case the two can be treated as one effective segment (see Fig.~\ref{fig:topos} (b)).\footnote{
The on-the-fly merging is also applied to open loops stemming from different Feynman diagrams, which have the same topology and last $N-k$ segments, but differ in the first $n$ segments,
after $k$ dressing steps, giving a speed-up of about a factor $2$ already without any on-the-fly reduction.}
The on-the-fly merging requires the multiplication of each Feynman diagram with its colour factor contracted with the full Born amplitude in the beginning. In fact, we 
initialize each diagram $d$ as
\be
\calU_0^{(d)}(\heli)=\calU_0^{(d)}(\{\heli_1,\ldots,\heli_N\})
=2\left(\sum_{\col}\calM^*_0(\heli)\,\calC_1^{(d)}\right),
\label{eq:initOL2}
\ee
and perform the dressing, reduction and merging steps on these extended open loops.
In this way diagrams with different original colour factors can be merged. The initial interference with the Born, which is computed for all global helicity configurations $\heli$ and
hence all helicity configurations $\heli_i$ of the individual segments, allows for an on-the-fly helicity summation, which significantly improves the CPU efficiency.
We define the partially dressed and helicity summed open loop as well as an extended dressing step
\bea
\calU_k(q,\{\heli_{k+1},\ldots,\heli_{N}\})
&=&\sum\limits_{\heli_{k}}\calU_{k-1}(q,\{\heli_k,\ldots,\heli_N\})\seg_k(q,\heli_k)
\eea
In this way, we reduce the number of dof of an open loop in the k-th dressing step by a factor equal to the number of helicity dof of the k-th segment. 
This new helicity treatment leads to an additional gain of a factor two--three in speed, depending on the process (see \cite{Buccioni:2017yxi}).

\section{Numerical stability} \label{sec:numstab}
In this section we present numerical stability studies for the on-the-fly algorithm\footnote{The final scalar integrals are evaluated with \Collier{}.} 
in \OpenLoops{} (OL2) compared to the previous version of \OpenLoops{} (OL1) using \Collier{} or \Cuttools{}
for the tensor integral reduction. The numerical accuracy of the double precision (DP) results is defined w.r.t.~a benchmark derived with OL1+\Cuttools{} in quadruple precision (QP),
$\calA = \log_{10} \left|(\calW_{\ssst{DP}}-\calW_{\ssst{QP}})/\min\left\{|\calW_{\ssst{DP}}|,|\calW_{\ssst{QP}}|\right\}\right|$.
To estimate the intrinsic accuracy of the QP benchmark we use a so-called rescaling test~\cite{Cascioli:2011va}. 

\begin{figure}[t!]\begin{center} \begin{tabular}{cc}
\includegraphics[width=0.50\textwidth]{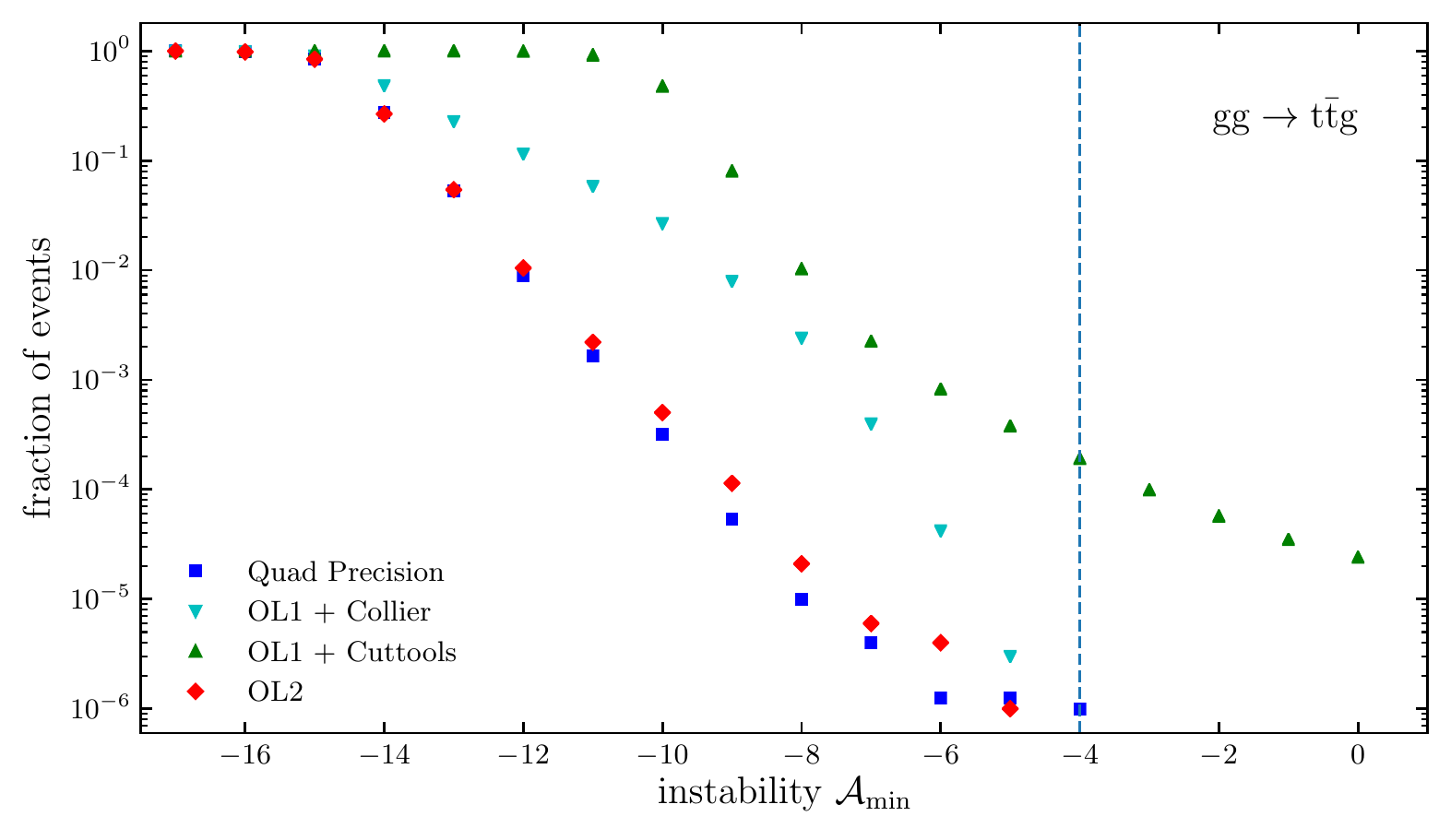} \hspace{-5mm} & 
\includegraphics[width=0.50\textwidth]{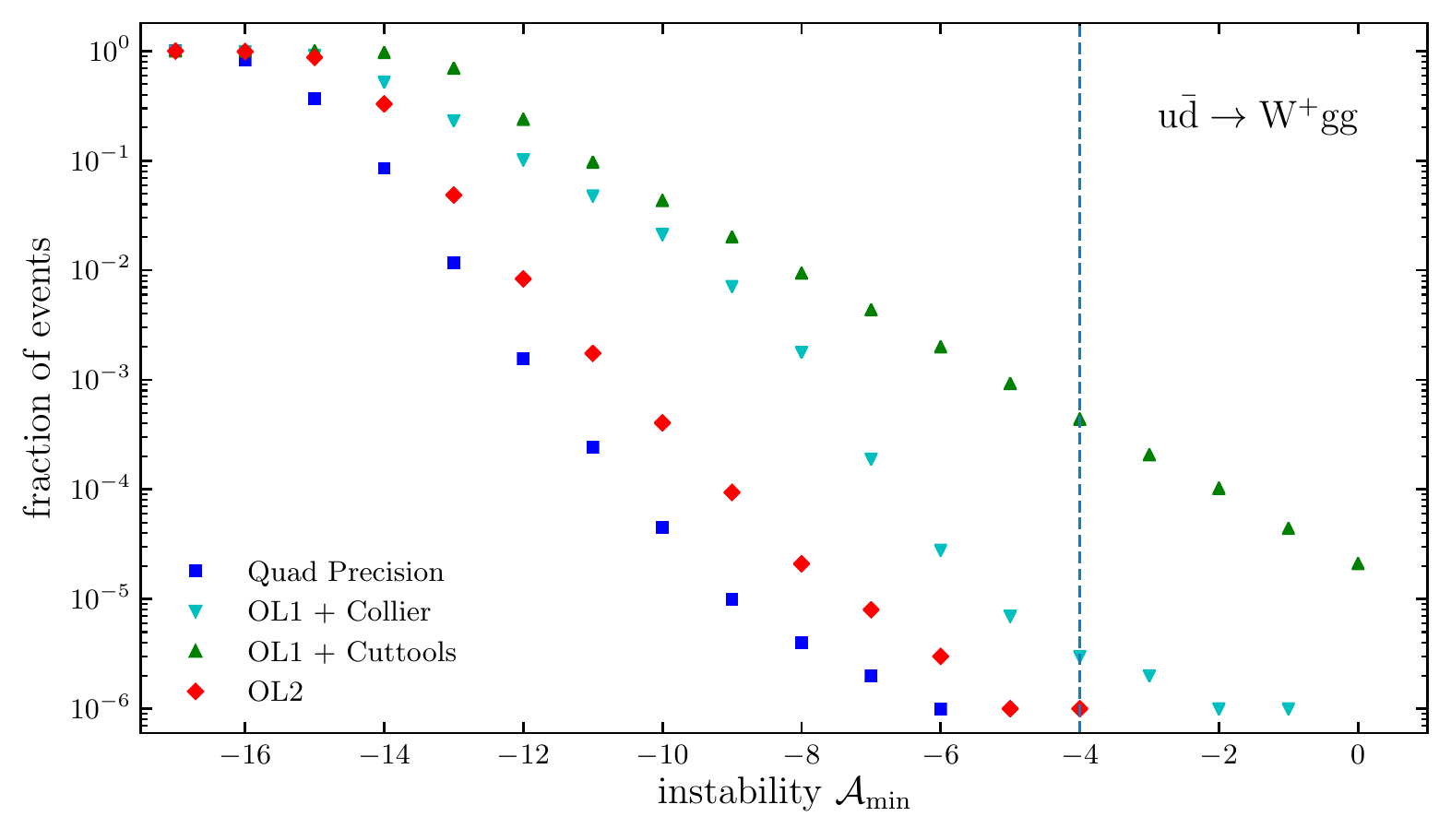}\\[-5mm]
  \end{tabular} \end{center}
\caption{Stability distributions for sample
$2 \to 3$  processes. \label{fig:numstab_2to3}}
\end{figure}

\begin{figure}[t!]\begin{center} \begin{tabular}{cc}
\includegraphics[width=0.50\textwidth]{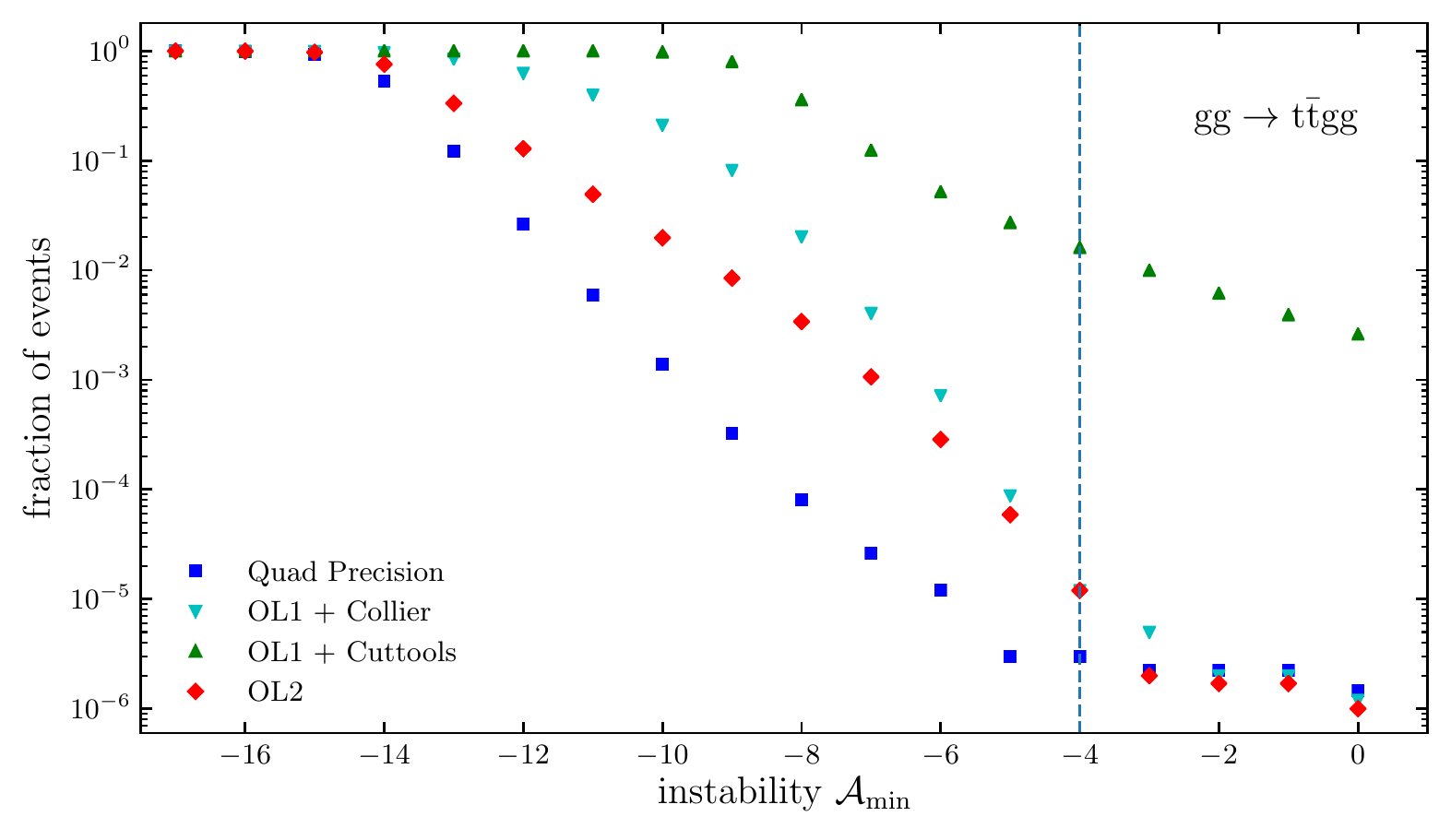} \hspace{-5mm} & 
\includegraphics[width=0.50\textwidth]{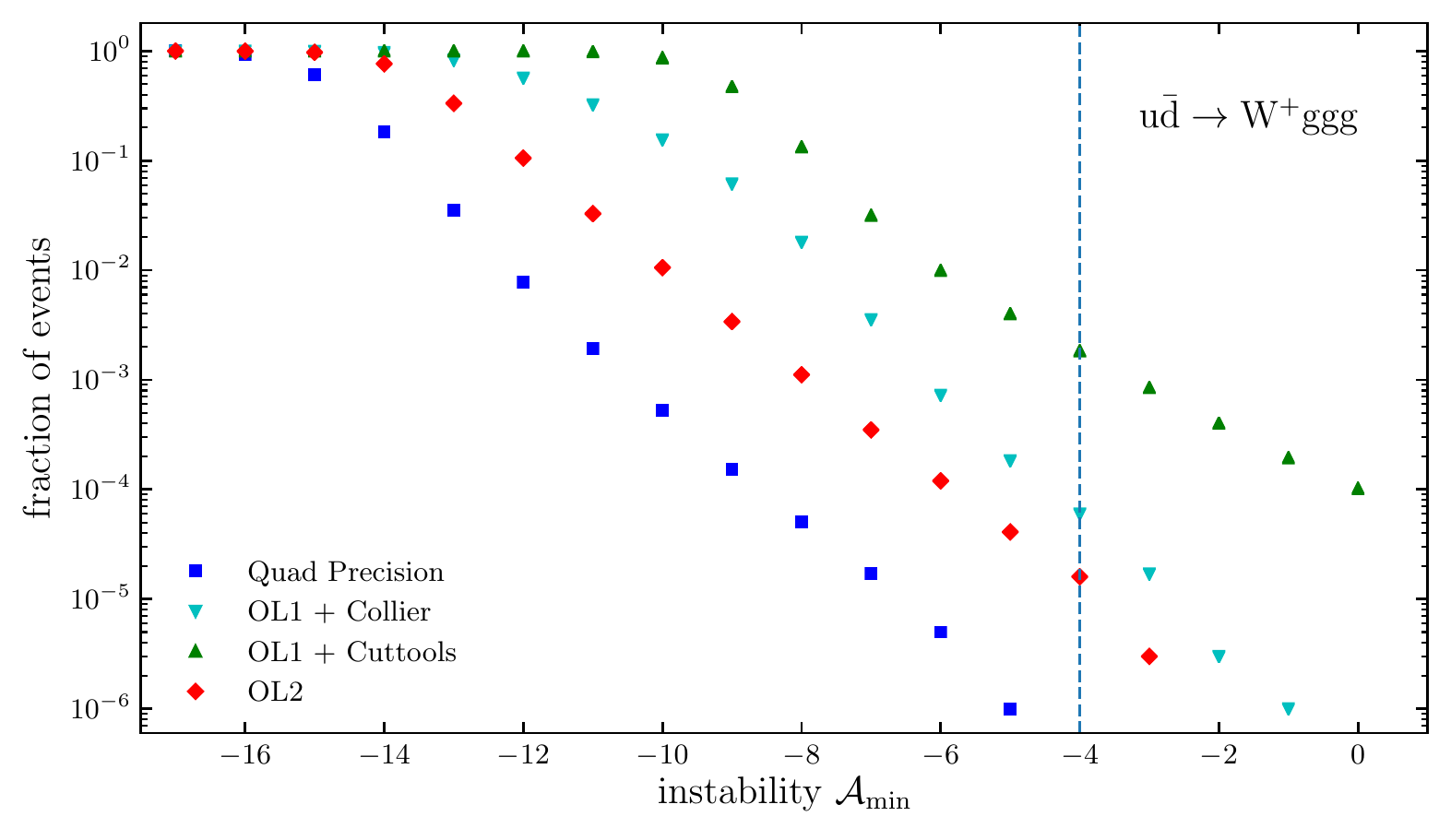}\\[-5mm]
  \end{tabular} \end{center}
\caption{Stability distributions for sample
$2 \to 4$  processes. \label{fig:numstab_2to4}}
\end{figure}

In Fig.~\ref{fig:numstab_2to3} and Fig.~\ref{fig:numstab_2to4} we show the fraction of points with an accuracy $\calA<\calA_{\ssst{min}}$ plotted against $\calA_{\ssst{min}}$ for sample
$2\to 3$ and $2\to 4$ processes respectively.\footnote{For each process a sample of $10^6$ homogeneously distributed
phase space points at $\sqrt{s}=1$\,TeV is taken. Infrared regions are excluded through cuts, $p_{i,\mathrm{T}} >
50$\,GeV and $\Delta R_{ij} > 0.5$, for massless final-state partons.}
While the results of OL1+\Cuttools{} feature the highest instability tails for
all considered processes, we find that using {\sc Collier} the probability of finding only a few correct digits goes down by
one to three orders of magnitude, depending on the process.
Using OL2 we observe improvements of one--two orders of magnitude w.r.t.~OL1+{\sc Collier} in many cases. For $2\to 3$ processes, 
the stability of the on-the-fly algorithm is remarkably close to the QP benchmark and even superior for $\mathrm{t\bar{t}g}$ production.  
In the case of $2\to 4$ processes OL1+{\sc Collier} and OL2 are very close in the tail, both achieving excellent numerical stability.

\section{Conclusion}

We have presented a new approach for the automated calculation of scattering
amplitudes at one loop.  The key idea is that various operations, such as tensor reduction, helicity summation and diagram merging, 
can be performed {\it on-the-fly} during the open-loop recursion, 
exploiting the factorised structure of open loops in a systematic way.
This reduces the complexity of certain operations in a very significant way.
The employed integrand reduction method allows us to isolate Gram determinant instabilities 
in triangle topologies with a particular kinematic configuration and to cure them by means of simple analytic expansions,
leading to an unprecedented level of numerical stability. This feature is particularly attractive
for the calculation of real--virtual contributions at NNLO.
The new algorithm is fully automated and validated at NLO QCD and will become publicly available in the upcoming release of {\sc OpenLoops\,2}.

\subsection*{Acknowledgments}
This research was supported in part by the Swiss National Science Foundation
(SNF) under contracts PP00P2-128552 and BSCGI0-157722.

\bibliographystyle{JHEP}

\providecommand{\href}[2]{#2}\begingroup\raggedright\endgroup

\end{document}